\documentclass[a4paper,12pt]{article}
\usepackage[pctex32]{graphics}
\usepackage{amssymb,amsmath}
\input epsf.sty
\newcommand{\be}{\begin{equation}}
\newcommand{\ee}{\end{equation}}
\newcommand{\ben}{\begin{eqnarray}\displaystyle}
\newcommand{\een}{\end{eqnarray}}

\begin{titlepage}

\begin{document}
{\baselineskip20pt


\vskip .6cm

\begin{center}
{\Large \bf New attractor mechanism for spherically symmetric
extremal black holes}

\end{center} }

\vskip .6cm
 \centerline{\large Yun Soo Myung$^{1,a}$,
 Yong-Wan Kim $^{1,b}$,
and Young-Jai Park$^{2,c}$}

\vskip .6cm

\begin{center}
{$^{1}$Institute of Mathematical Science and School of Computer
Aided Science, \\Inje University, Gimhae 621-749, Korea \\}

{$^{2}$Department of Physics and Center for Quantum Spacetime,
Sogang University, Seoul 121-742, Korea}
\end{center}

\vspace{5mm}

\centerline{\bf Abstract} \bigskip

We introduce a new attractor mechanism to find the entropy for
spherically symmetric extremal black holes. The key ingredient  is
to find a two-dimensional (2D) dilaton gravity with the dilaton
potential $V(\phi)$. The condition of  an attractor is given by
$\nabla^2\phi=V(\phi_0)$ and $\bar{R}_2=-V^{\prime}(\phi_0)$ and for
a constant dilaton $ \phi=\phi_0$, these are also used to find the
location of the degenerate horizon $r=r_{e}$ of an extremal black
hole. As a nontrivial example, we consider an extremal regular black
hole obtained  from  the coupled system of Einstein gravity and
nonlinear electrodynamics. The desired Bekenstein-Hawking entropy is
successfully recovered from the generalized entropy formula combined
with the 2D dilaton gravity, while the entropy function approach
does not work for obtaining this entropy.

\vskip 0.6cm



\vskip 0.8cm

\noindent $^a$ysmyung@inje.ac.kr \\
\noindent $^b$ywkim65@gmail.com \\
\noindent $^c$yjpark@sogang.ac.kr

\noindent
\end{titlepage}

 \setcounter{page}{2}

\section{Introduction}

Still growing interest in extremal  black holes is motivated by
their unusual and not fully understood nature. The problems of
entropy,  semiclassical configurations, interactions with matter,
and  information paradox have not been resolved completely. Apart
from their global structure and behavior, the near horizon region is
also of interest~\cite{gp}.

In particular, of the equal importance is the question of the
singularities that reside in centers of most black holes hidden to
an external observer. Regular black holes (RBHs) have been
considered, dating back to Bardeen~\cite{BAR}, for avoiding the
curvature singularity beyond the event horizon~\cite{RBH}. Their
causal structures are similar to a Reissner-Nordstr\"{o}m (RN) black
hole with the singularity replaced by de Sitter
space-time~\cite{Dymn1}. In addition to various RBHs~\cite{HAY}, the
action of Einstein gravity and nonlinear electrodynamics provided a
magnetically charged RBH~\cite{0006014}. This solution is featured
by two integration constants and a free parameter. The integration
constants are related to Arnowitt-Deser-Misner (ADM) mass $M$ and a
magnetic charge $Q$, while the free parameter $a$ is adjusted to
make the line element regular at center. Moreover, it allows exact
treatment by using the Lambert function~\cite{0010097}. Here we note
that this extremal RBH has the near horizon geometry of $AdS_2\times
S^2$ as the extremal RN black hole does have~\cite{0403109,0606185}.

On the other hand, string theory suggests that higher curvature
terms can be added to Einstein gravity~\cite{Zwiebach}. Black holes
in higher-curvature gravity \cite{Callan} were extensively studied,
showing the spectacular progress in the microscopic counting of
black hole entropy. For a review, see \cite{deWit}. In theories with
higher curvature corrections, classical entropy deviates from the
Bekenstein-Hawking  entropy  and can be calculated using Wald's
Noether charge formalism \cite{Wald}. It exhibits exact agreement
with string theory predictions, both in the BPS \cite{Behrndt} and
non-BPS \cite{Goldstein,AE} cases.

Recently, Sen has proposed a so-called ``entropy function'' method
for calculating the entropy of $n$-dimensional extremal black holes,
which is  effective even for the presence of higher curvature terms.
Here the extremal black holes are characterized by the near horizon
geometry of $AdS_{2}\times S^{n-2}$ and corresponding
isometry~\cite{Sen}. It states that the entropy of extremal black
holes can be obtained by extremizing the entropy function with
respect to some moduli on the horizon. Extremizing a entropy
function is equivalent to solving Einstein equation in the near
horizon. A entropy function usually depends only on the near horizon
geometry, and decouples from  the data at infinity. This describes
the attractor behavior. This method  has been applied to many
solutions including extremal black holes in higher dimensions,
rotating black holes and various non-supersymmetric black
holes~\cite{SSen}. We note that the near horizon isometry SO(2,1)
and the long throat of $AdS_2$ sector are the two ingredients of the
attractor mechanism~\cite{klr}. On the other hand, Cai and
Cao~\cite{CC} have proposed the generalized entropy formula based on
Wald's Noether charge formalism. However, this generalized entropy
function approach has a drawback that there is no way to combine the
full equations of motion with  the attractor mechanism.

Very recently, we have investigated a magnetically charged
RBH~\cite{MKP}. It turned out that the entropy function approach
does not work for deriving  the Bekenstein-Hawking entropy of the
extremal RBH, while the generalized entropy formula is suitable
for the RBH case. This is mainly because the  entropy function
depends on  the near horizon geometry ($Q^2$) nonlinearly as well
as  the data at infinity ($M$).

In this paper, we address this issue of the extremal RBH again. We
introduce a new attractor mechanism to find Bekenstein-Hawking
entropy for the extremal RBH.  The important point is to find a 2D
dilaton gravity with dilaton potential $V(\phi)$ by imposing the
dimensional reduction of 4D Einstein gravity including matter and
then by performing a conformal transformation \cite{NSN,FNN}.
Then, the  new attractor equations  are given by
$\nabla^2\phi=V(\phi)$ and $V^{\prime}(\phi)=-\bar{R}_2$.  For a
constant dilaton $\phi=\phi_0$, this can be used to find the
location $r=r_{e}$ of degenerate horizon. Finally, we use the
generalized entropy formula based on Wald's Noether charge
formalism to derive the desired Bekenstein-Hawking entropy.

The organization of this work is as follows. In Sec. 2, we show
the procedure to find a 2D dilation gravity from a 4D Einstein
gravity coupled with matter. In order to test whether this
approach works for calculating the entropy of an extremal black
hole, we study a toy model of the RN black hole in Sec. 3. We
briefly review  a magnetically charged RBH in Sec. 4. In Sec. 5,
we show that Sen's entropy function approach does not work for the
regular black hole. Sec. 6 is devoted to finding the entropy of an
extremal RBH by using the generalized entropy function approach.
We obtain  the Bekenstein-Hawking entropy of an extremal RBH by
using  the 2D dilaton gravity approach with a conformal
transformation in Sec. 7. Finally, we discuss our results in Sec.
8.

\section{Dimensional Reduction Approach}
We start with the four-dimensional (4D) action
\begin{equation}
\label{action}
I=\frac{1}{16\pi}\int d^4x \sqrt{-g}[R-{\cal L}_M(B)]
\end{equation}
where ${\cal L}_M(B)$ is the Lagrangian for matter. For our purpose,
we consider  the spherically symmetric metric
\begin{equation}
\label{metric}
 ds^2=-U(r)dt^2+\frac{1}{U(r)}dr^2+ b^2(r) d\Omega^2_2,
\end{equation}
where $b(r)$ plays a role of the radius of two sphere $S^2$. The
4D Ricci scalar $R$ is calculated as
\begin{equation}
R=-U''-\frac{1}{b^2}\Big[4b b' U'+2U (b'^2+2b b'')-2\Big],
\end{equation}
where the prime denotes the derivative with respect to $r$. After
the dimensional reduction by integrating the action (\ref{action})
over $S^2$, the reduced effective action in two
dimensions~\cite{NSN} can be rewritten by
\begin{equation}
I^{(2)}=\frac{1}{4}\int d^2x\sqrt{-g}
     [b^2 R_2+2g^{\mu\nu}\nabla_\mu b\nabla_\nu b+2-4 b^2{\cal
     L}_M],
\end{equation}
where $R_2=-U''(r)$ is the 2D Ricci scalar. It is convenient to
eliminate the kinetic term by using the conformal transformation
\begin{equation} \label{conft}
\bar{g}_{\mu\nu}=\sqrt{\phi}~g_{\mu\nu},~\phi=\frac{b^2(r)}{4}.
\end{equation}
This transformation delivers information on the 4D action
(\ref{action}) to 2D dilaton potential, if the 4D action provides
the black hole solution.  That is, we may get the good $s$-wave
approximation to the 4D black hole eliminating the kinetic term.
Unless one makes the conformal transformation, the information is
split into the kinetic and the potential terms.

Now, let us choose the dilaton as the squared radius of $S^2$
($\phi=r^2/4$). Then, the reparameterized action takes the form
\begin{equation}\label{repara-action}
\bar{I}^{(2)}=\int d^2x \sqrt{-\bar{g}}[\phi\bar{R}_2+V(\phi)]
             \equiv\int d^2x\sqrt{-\bar{g}}\bar{F},
\end{equation}
where the Ricci scalar and the dilaton potential are given by
\begin{equation}
\bar{R}_2=-\frac{U''}{\sqrt{\phi}},~V(\phi)=\frac{1}{2\sqrt{\phi}}-\sqrt{\phi}{\cal
L}_M(B),
\end{equation}
respectively. This is a 2D dilaton gravity (for a review, see
~\cite{gkv}) with $G_2=1/2$~\cite{JT}. Also $\bar{F}$ will play a
role of the entropy function. The two equations of motion are
derived as
\begin{eqnarray} \label{newatt1}
\nabla^2\phi=V(\phi),\\
\bar{R}_2=-V'(\phi),  \label{newatt2}
\end{eqnarray}
where $V'(\phi)$ denotes the derivative with respect to $\phi$.
These equations will play the role of  a new attractor. Note that
in the case of nonsupersymmetric attractors in four
dimensions~\cite{Goldstein}, the condition for attractor are that
$\partial_iV_{eff}(\phi_{i0})=0$ and
$\partial_i\partial_jV_{eff}(\phi_{i0}) $ should have positive
eigenvalues at the critical point of $\phi_i=\phi_{i0}$. If these
conditions are satisfied, the attractor mechanism works and the
entropy is given by the effective potential at horizon.

For our case, the corresponding conditions for the  attractor
 are
\begin{equation}\label{attcon}
V(\phi_0)=0, ~~~V'(\phi_0)\neq 0
\end{equation}
for  the constant dilaton $\phi= \phi_0$. A solution to the
equation (\ref{newatt1}) provides a constant dilation $\phi_0$
when $V(\phi_0)=0$. Considering the connection of
$\phi_0=r_e^2/4$,  the solution to the equation (\ref{newatt2})
gives us information on the 2D spacetime
\begin{equation}
\bar{R}_2|_{r=r_e}=-\frac{U''(r_e)}{\sqrt{\phi_0}}
\end{equation}
which is  a constant curvature of the $AdS_2$ spacetime.

After the conformal transformation,  the generalized entropy formula
derived by Wald's Noether charge formalism is obtained as
\begin{equation}
\bar{S}_{BH}=\frac{4\pi \sqrt{\phi_0}}{U''(r_e)}(q e - \bar{F}),
\end{equation}
which is slightly  different from the entropy formula
$S_{BH}=\frac{4\pi }{U''(r_e)}(q e - F)$, proposed by Cai and Cao
\cite{CC}. Finally,  we have the generalized entropy function
\begin{equation}
\bar{F}(\phi_0)= - \sqrt{\phi_0} U''(r_e)
\end{equation}
at the degenerate horizon.  Considering $\phi_0=
\frac{1}{4}r_e^2$, the desired Bekenstein-Hawking entropy is
obtained by
\begin{equation}
\bar{S}_{BH}=-\frac{4\pi\sqrt{\phi_0}}{U''(r_e)}\bar{F}(\phi_0)=4\pi\phi_0
            = \pi r_e^2
\end{equation}
for a magnetically charged black hole with $e=0$.

\section{Reissner-Norstr\"om black hole}

We consider a toy model of Einstein-Maxwell theory to test whether
our approach does work for obtaining a proper entropy of extremal RN
black hole. In this case, we have
\begin{equation}
{\cal L}_M(B)=F_{\mu\nu}F^{\mu\nu}=\frac{2Q^2}{r^4}
\end{equation}
with $F_{\theta\varphi}=Q\sin\theta$. Then, the potential is given
by
\begin{equation}
V(\phi)=\frac{1}{2\sqrt{\phi}}\Big[1-\frac{Q^2}{4\phi}\Big]
\end{equation}
whose form is depicted in Fig. 1. When $V(\phi_0)=0$, one has the
solution to Eq. (\ref{newatt1}) as
\begin{equation} \phi_0=\frac{Q^2}{4}.
\end{equation}
In this case, we have the $AdS_2$ spacetime  with the curvature
\begin{equation}
\bar{R}_2|_{r=r_e}=-V'(\phi_0)=-4/Q^3.
\end{equation} Finally, since the
generalized entropy function is given by
\begin{equation} \bar{F}^{RN}(\phi_0)= -\sqrt{\phi_0}
U''(r_e),
\end{equation}
we have the entropy for the extremal RN black hole as
\begin{equation}
\bar{S}^{RN}_{BH} = 4\pi\phi_0=\pi Q^2.
\end{equation}
As is shown in Fig. 1, one cannot find the degenerate horizon for
$Q^2=0$ case because it corresponds to Schwarzschild black hole.
\begin{figure}[t!]
   \centering
   \includegraphics{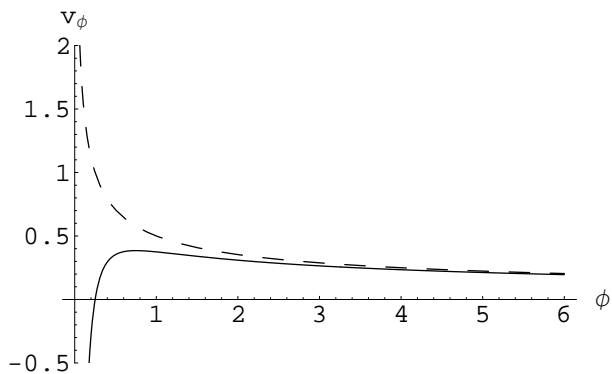}
\caption{The solid curve: the dilaton potential $V(\phi)$ for the
RN black hole with $Q=1$. $V(\phi)=0$ at $\phi=\phi_0=0.25$
denotes the degenerate horizon. For $Q\not=0$, one always finds
the point of $V(\phi)=0$. The large-dashed curve is for the
Schwarzschild case with $Q^2=0$, where there is no point of
$V(\phi)=0$.} \label{fig1}
\end{figure}

\section{Regular black hole}

We briefly review a magnetically charged
RBH~\cite{0403109,0606185}. In this case, ${\cal L}_M(B)$ in Eq.
(\ref{action}) is a functional of $B= F_{\mu\nu}F^{\mu\nu} $
defined by
\begin{equation}  \label{lagr}
 {\cal L}_M(B)=B \cosh^{-2}\left[a \left(\frac{B}{2}\right)^{1/4} \right],
 \end{equation}
where the free parameter $a$ will be adjusted to guarantee the
regularity at the center. In the limit of $a\to 0$, we recover the
Einstein-Maxwell theory in the previous section. To determine the
metric function (\ref{metric}) defined  by
\begin{equation}
U(r)\,=\,1\,-\,\frac{2 m(r)}{r},
 \end{equation}
we have to solve Einstein equation. From the variation of the
action (\ref{action}) together with the matter (\ref{lagr}) with
respect to the vector potential  $A_\mu$, the equations of motion
are given by
\begin{equation}
\label{maxwell1} \nabla _{\mu}\left( \frac{d{\cal L}(B)}{dB}
F^{\mu\nu}\right) =0,
\end{equation}
\begin{equation}
\label{maxwell2} \nabla _{\mu}\,^{\ast }F^{\mu\nu}=0,
\end{equation}
where the asterisk denotes the Hodge duality. The solution to Eqs.
(\ref{maxwell1}) and (\ref{maxwell2}) is $F_{\theta\varphi}=Q\sin
\theta$ for  a magnetically charged case. On the other hand, the
variation of the action with respect to the metric $g_{\mu\nu}$
leads to the Einstein equation
\begin{equation} \label{eineq}
R_{\mu\nu}-\frac{1}{2} g_{\mu\nu}R = 8\pi T_{\mu\nu}
\end{equation}
with the stress-energy tensor
\begin{equation}
T_{\mu\nu}=\frac{1}{4\pi }\left( \frac{d {\cal L}\left( B\right)
}{dB}F_{\rho \mu}F^{\rho}_{\nu}-\frac{1}{4}g_{\mu\nu} {\cal L}\left(
B\right) \right).
\end{equation}
After solving these Einstein equation, the mass distribution is
determined to be
\begin{equation} \label{quadrature1}
m(r)\,=\,\frac{1}{4}\int^r  {\cal L}[B(r')]r'^{2} dr'\, + C,
\end{equation}
where $C$ is an integration constant. Considering  the condition for
the ADM mass $M(=m(\infty)=C)$, the mass distribution takes the form
\begin{equation}
m(r)\,=M-\frac{Q^{3/2}}{2a} \tanh\left(\frac{aQ^{1/2}}{r} \right).
\end{equation}
Moreover, setting $a\,=\,Q^{3/2}/2M$ determines the metric function
completely as
\begin{equation}
U(r)\,=\,1\,-\,\frac{2 M}{r}\left(1\,-\,\tanh\frac{Q^{2}}{2Mr}
\right). \label{Gr}
 \end{equation}
At this stage we note that $U(r)$ is regular as $r \to 0$, in
contrast to the RN case ($a=0$ limit) where its metric function of
$1-2M/r+Q^2/r^2$ diverges as $r^{-2}$ in that limit.  In order to
find the location of the horizon from $U(r)=0$, we use the
 Lambert functions $W_i (\xi)$ defined by the
formula $W(\xi)e^{W(\xi)}=\xi$ \cite{0403109}. As is shown in Fig.
2, $W_0(\xi)$ and $W_{-1}(\xi)$ are real branches. Their values at
branch point $\xi=-1/e$ are the same as
$W_{0}(-1/e)=W_{-1}(-1/e)=-1$.
 Here we set
$W_{0}(1/e) \equiv w_0$ because  it  plays an important role in
finding the location of degenerate horizon of the extremal RBH.
\begin{figure}[t!]
   \centering
   \includegraphics{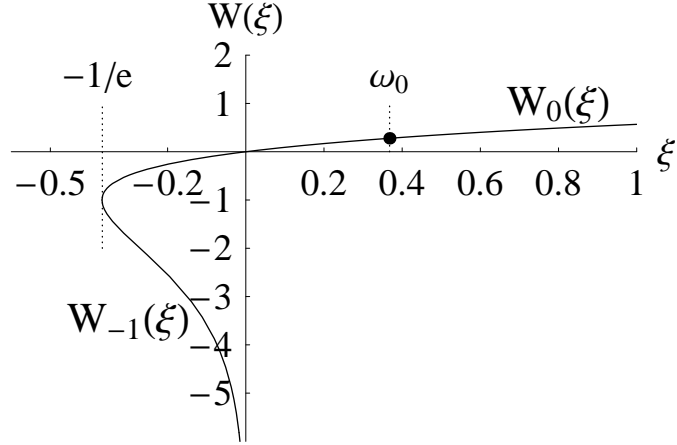}
\caption{The two real branches of the Lambert function $W_0(\xi)$
(upper curve) and $W_{-1}(\xi)$ (lower curve) are depicted for the
solution to the RBH. The degenerate event horizon at ($q_e,x_e$)
corresponds to the branch point of the Lambert function at
$\xi=-1/e$. } \label{fig2}
\end{figure}
We note that the mass $M$ is a free parameter.  Introducing  a
reduced radial coordinate $x=r/M$ and a charge-to-mass ratio
$q=Q/M$, the condition for the event horizon is given by
\begin{equation}
U(x(q))\,=\,1\,-\,\frac{2 }{x}\left(1\,-\,\tanh\frac{q^{2}}{2x}
\right)=0. \label{Grn}
 \end{equation}
Here one finds the outer $x_+$ and inner $x_-$ horizons as
\begin{equation}
x_+(q)=-\frac{q^2}{W_0(-\frac{q^2e^{q^2/4}}{4})-q^2/4},
~~x_-(q)=-\frac{q^2}{W_{-1}(-\frac{q^2e^{q^2/4}}{4})-q^2/4}.
\end{equation}
For $q^2/4=q^2_{e}/4=w_0$, the two horizons $x_+$ and $x_-$ merge
into a degenerate event horizon\footnote{The near horizon geometry
  of the degenerate horizon
  $U(r) \simeq h(r-r_{e})^2$ with $U'(r_{e})=0$ and
  $U''(r_{e})=2h$. Introducing new coordinates
  $r= r_{e} + \varepsilon/(hy$) and $\tilde{t}=t/\varepsilon$
  with $h=\frac{(1+\omega_o)^3}{32M^2 \omega_{o}^2}.$
  Expanding the function $U(r)$ in terms of $\varepsilon$,
  retaining quadratic terms and subsequently taking the limit of
  $\varepsilon \to 0$, the line element~\cite{0403109} becomes
  $ds^2_{NH} \simeq \frac{1}{hy^2} \left( -dt^2 + dy^2 \right) +
  r^2_{e} d\Omega_2^2.$ Moreover, using the Poincar\`{e}
  coordinate $y =1/u$, one could rewrite the above line element as
  the standard form of $AdS_2\times S^2$: $ ds^2_{NH} \simeq
  \frac{1}{h} \left( - u^2 dt^2 + \frac{1}{u^2}du^2 \right) +
  r^2_{e} d\Omega_2^2$.}
 at
\begin{equation}
x_{e}=\frac{4q^2_{e}}{4+ q^2_{e}}=\frac{4w_0}{1+w_0},
\end{equation}
where we use the relation of
$(q_{e}^2/4)e^{q_{e}^2/4}=1/e=w_0e^{w_0}$. That is, the degenerate
event horizon appears  at $(q_{e}=1.056, x_{e}=0.871)$ when
$x_+=x_-=x_e$. We note that in finding the location of the
degenerate horizon, first we choose $q=q_e$ and then determine
$x=x_e$.  For $q>q_{e}$, there is no horizon, while for $q<q_e$, two
horizons appear. For our purpose, let us define the
Bekenstein-Hawking entropy for the magnetically charged extremal RBH
as
\begin{equation} \label{BHRBH}
S_{BH}=  \pi r^2_{e}
  =\pi M^2 x^2_{e}=\pi Q_e^2\Big[\frac{4 q_{e}}{4+ q^2_{e}}\Big]^2
\end{equation}
with $Q_e=Mq_e$.

\section{Entropy function approach}

The magnetically charged extremal RBH is an interesting object
beacuse its near horizon geometry is given by the topology of
$AdS_2\times S^2$ and its action is already known.  Let us attempt
to derive  the black hole entropy in Eq. (\ref{BHRBH}) using Sen's
entropy function approach. For this purpose, we consider an extremal
black hole solution whose near horizon geometry is given by
$AdS_2\times S^2$ with the magnetically charged configuration
\begin{eqnarray}
\label{e11}
 && ds^2\equiv g_{\mu\nu}dx^\mu dx^\nu =
    v_1\left(-r^2 dt^2+{dr^2\over r^2}\right) +  v_2~ d \Omega^2_2,  \\
\label{e12}
 &&  F_{\theta\phi} = {Q} \, \sin\theta\, ,
\end{eqnarray}
where $v_i (i=1,2)$ are constants to be determined. Now, let us
define the Lagrangian density $f(v_i, Q)$ as the remaining part
after integrating the action (\ref{action}) over $S^2$ as
follows~\cite{0505122}:
\begin{equation}
\label{e2} {f}(v_i, Q) = \frac{1}{16\pi} \int d\theta\, d\phi\,
\sqrt{- g}\, \left[R\, -\,{\cal L}_M(B) \right]\,.
\end{equation}
Since $ R=-\frac{2}{v_1}+\frac{2}{v_2}$ and
$B=\frac{2{Q}^2}{{v_2}^2}$, we obtain
\begin{equation}
{f}(v_i, {Q})=\frac{1}{2}v_{1}v_2\left[-\frac{1}{v_1}+\frac{1}{v_2}
\, - \frac{Q^2}{v^2_2}\cosh^{-2}\Big(\frac{Q^2}{2M\sqrt{v_2}}\Big)
\right].
\end{equation}
Here, we choose the free parameter $a=Q^{3/2}/2M$ to have a RBH
solution.
 For the magnetically charged extremal RBH, the entropy function is given by
\begin{equation} \label{e31}
{\cal F}(v_i, {Q}) = -2\pi {f}(v_i, {Q}) \, .
 \end{equation}
In this case, the extremal values $v_i^e$  are determined by
extremizing the function ${\cal F}(v_i, {Q}) $ with respect to
$v_i$:
\begin{eqnarray}
\label{e33} \frac{\partial {\cal F}}{\partial v_1}&=&0 \Rightarrow v_2=Q^2\cosh^{-2}
\left[\frac{Q^2}{2M\sqrt{v_2}}\right], \\
\frac{\partial {\cal F}}{\partial v_2}&=&0 \Rightarrow
\frac{1}{v_1}=\frac{Q^2}{v^2_2}\cosh^{-2}\left[\frac{Q^2}{2M\sqrt{v_2}}\right]
 -\frac{Q^2}{v_2}\frac{\partial}{\partial
v_2}\left(\cosh^{-2}\left[\frac{Q^2}{2M\sqrt{v_2}}\right]\right).\nonumber\\
\label{e332}
\end{eqnarray}
Actually, these are conventional attractor equations. Using the
above relations,  the entropy function at the extremum is given by
\begin{equation}
{\cal F}(v^e_{2}, Q)=\pi v^e_2.
 \end{equation}
In order to find the  extremal value of $v^e_2$, we introduce $Q =
\tilde{q} M$ and  $v^e_2 = M^2 \tilde{x}^2 $. Then, Eqs. (\ref{e33})
and (\ref{e332})
 can be rewritten as the following  equations
\begin{eqnarray}
\label{sol1}\frac{\tilde{x}^2}{\tilde{q}^2}&=&\cosh^{-2}(\frac{\tilde{q}^2}{2\tilde{x}}), \\
\label{sol2}\frac{1}{v_1}&=& \frac{\tilde{q}^2}{M^2{\tilde{x}}^4}
\cosh^{-2}(\tilde{q}^2/2\tilde{x})
-\frac{\tilde{q}^4}{2M^2\tilde{x}^5}
\frac{\sinh(\tilde{q}^2/2\tilde{x})}{\cosh^{3}(\tilde{q}^2/2\tilde{x})},
\end{eqnarray}
which  are identical to the  Einstein equation in the near horizon
geometry of the $AdS_2\times S^2$ in Ref.{~\cite{0403109}} except
being replaced $1/v_1$ by $h$. This means that the entropy function
approach is equivalent to solving the Einstein equation on the
$AdS_2 \times S^2$ background (the attractor equations). Of course,
Eqs. (\ref{e33}) and (\ref{e332}) are not the full Einstein equation
in (\ref{eineq}).
\begin{figure}[t!]
   \centering
   \includegraphics{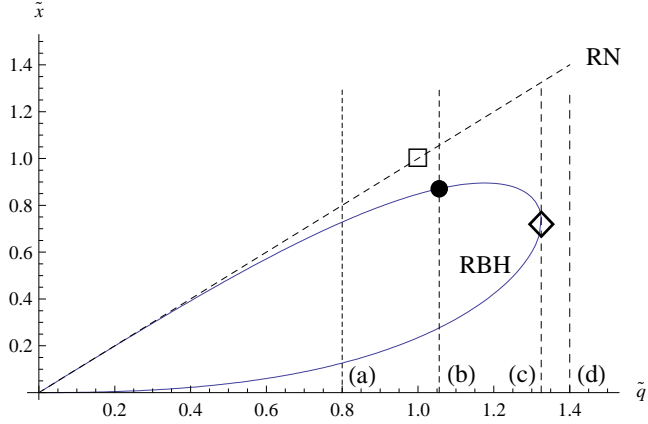}
\caption{Plot of the curvature radius $\tilde{x}$ of $S^2$ versus
the parameter $\tilde{q}$. The two  curves  denote  the solution
space of the near horizon geometry to Eq. (\ref{sol1}), while the
line denotes $\tilde{x}=\tilde{q}$ for the RN case with its
extremal point $\tilde{x}=\tilde{q}=1(\square)$. The upper curve
includes a point of $\bullet$, which corresponds to an extremal
RBH. However, the lower curve belongs to the unphysical solution
space because of negative $v_1$. $\diamond$ denotes the critical
point at ($\tilde{q}_c,\tilde{x}_c$). (a)-(d) are introduced to
connect the 2D dilaton potential in Fig. 4.}\label{fig.3}
\end{figure}

Since the above equations are nonlinear differential equations, we
could not solve them analytically. In the limit of $a \to 0$, one
easily finds the extremal RN case such that
$v^e_2=M^2\tilde{q}^2(\tilde{x}=\tilde{q})$, $v_1^e=v_2^e=Q^2$.
Instead, we have to solve numerically the Eq. (\ref{sol1}) because
of the nonlinearity between $\tilde{x}$ and $\tilde{q}$. Their
solutions are depicted in Fig. 3. In this figure, the solid line
corresponds to the solution space in which each set of $(\tilde{q},
\tilde{x})$ resides on the subspace $S^2$. There are two branches:
the upper and lower ones which merge at the critical point of  $
(\tilde{q}_c,\tilde{x}_c)=(1.325,0.735)$. Since the lower branch
gives us negative $v_1$ and thus it becomes de Sitter space instead
of $AdS_2$, this branch should be ruled out. Note that  the
magnetically charged extremal RBH corresponds to the point
$(\tilde{q}_{e},\tilde{x}_{e})=(1.056, 0.871)$. However, there is no
way to arrive at this point even though the solution space comprises
such a point. In this case,  the entropy function takes the form
\begin{equation}\label{sen1}
{\cal F}= \pi v^e_2= \pi M^2 \tilde{x}^2 =\pi M^2\tilde{q}^2
\cosh^{-2}\left(\frac{\tilde{q}^2}{2\tilde{x}}\right).
\end{equation}
 We note that this entropy function depends on both $\tilde{q}$
and $\tilde{x}$, in contrast to the RN case of ${\cal F}_{RN}=\pi
M^2 \tilde {x}^2=\pi Q^2$. Hence, unless one knows $\tilde{q}=q_e$
and $\tilde{x}=x_e$, we cannot obtain the Bekenstein-Hawking entropy
of $S_{BH}=\pi M^2x_{e}^2$ in Eq. (\ref{BHRBH}).

\section{Generalized entropy function approach}

Before performing the conformal transformation, the entropy formula
based on Wald's Noether charge formalism \cite{CC} takes the form
\begin{equation} \label{WEF}
S_{BH}=\frac{4\pi }{U''(r_e)}\left(q e - F(r_e)\right),
\end{equation}
where the generalized entropy function $F$ is given by
\begin{equation}
F(r_e)= \frac{1}{16\pi}\int_{r=r_e} d\theta d\varphi
\sqrt{-g}\left[R-{\cal L}_M\right]
\end{equation}
with
\begin{eqnarray}
&& R=-\frac{r^2U''+4 r U'+2U -2}{r^2},\\
 && {\cal L}_M=\frac{2Q^2}{r^4}\cosh^{-2}\left[\frac{Q^2}{2Mr}\right].
\end{eqnarray}
In this approach, one has to know the location of the degenerate
horizon (the solution to full Einstein equations). After the
integration over the angular coordinates, the generalized entropy
function leads to
\begin{equation}
F(r_e)= \left.\frac{1}{4}\Big[-r^2U''(r)+2-r^2{\cal
L}_M\Big]\right|_{r=r_e}=-\frac{1}{4}U''(r_e)r_e^2
\end{equation}
because of ${\cal L}_M \mid_{r=r_e}=\frac{2}{r^2_e}$ and
$U(r_e)=U'(r_e)=0$.
 Finally, for $e=0$ we obtain the correct form of the entropy from
 Eq. (\ref{WEF}) as
\begin{equation} \label{WEFe1}
S_{BH}=-\frac{4\pi }{U''(r_e)}F(r_e)=\pi r_e^2.
\end{equation}
Even though we find the Bekenstein-Hawking entropy using the
generalized entropy formula, there is still  no way to fix the
location $r=r_e$ of the degenerate horizon. Hence we have to find
another approach to calculate the entropy of an extremal RBH
natually.

\section{2D dilaton gravity  approach}

Now, the remaining issue is how to incorporate  the full equations
of motion to extremizing process to find the entropy of an extremal
RBH. We start with the action of the 2D dilation gravity in Eq.
(\ref{repara-action})
\begin{equation}\label{JTA}
\bar{I}_{RBH}=\int d^2x
\sqrt{-\bar{g}}\left[\phi\bar{R}_2+V(\phi)\right]=\int d^2x
\sqrt{-\bar{g}} \bar{F},
\end{equation}
where  the Ricci scalar and the dilaton potential are
\begin{equation}
\bar{R}_2=-\frac{U''}{\sqrt{\phi}},~V(\phi)=\frac{1}{2\sqrt{\phi}}-
\frac{Q^2}{8\phi^{3/2}}\cosh^{-2}\left[\frac{Q^2}{4M\sqrt{\phi}}\right],
\end{equation}
respectively.  The two equations of motion are obtained as
\begin{eqnarray}
\label{tee33} \nabla^2\phi&=& V(\phi), \\
\bar{R}_2&=&-V'(\phi) \label{tee34},
\end{eqnarray}
which give the equations for a new attractor. The solution to
these equations provides the ground state for the $AdS_2$-gravity
of Jackiw-Teitelboim theory \cite{JT}. Without any gauge-fixing,
the solution to Eq. (\ref{tee33}) may be a constant dilation
$\phi_0$ when $V(\phi_0)=0$.  $V(\phi_0)=0$ implies
\begin{equation}
\label{tee44}
\phi_0=\frac{Q^2}{4}\frac{1}{\cosh^{2}[\frac{Q^2}{4M\sqrt{\phi_0}}]}.
\end{equation}
\begin{figure}[t!]
   \centering
   \includegraphics{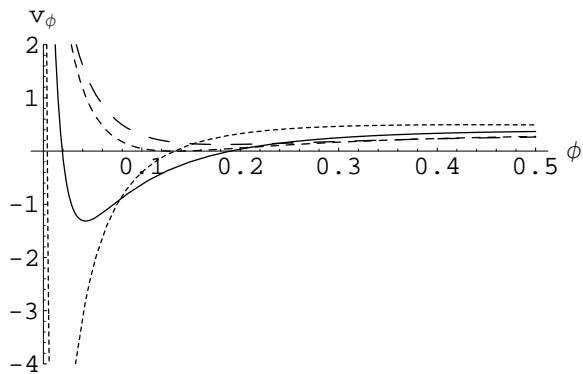}
\caption{The solid curve: the dilaton potential $V(\phi)$ for the
extremal RBH with $M=1,Q_e=1.056$((b) in Fig. 3). $V(\phi_0)=0$ is
at $\phi_0$ = 0.019(unphysical) and $\phi_0 = 0.19(r_e=0.87)$ where
the latter denotes the degenerate horizon. The large-dashed curve is
for no horizon with $Q=1.4$ ((d) in Fig. 3) where there is no point
of $V(\phi)=0$. The small-dashed curve is for the critical case with
$Q=1.325$((c) in Fig. 3), which implies one point of $V(\phi_0)=0$.
The dotted curve is for another extremal black hole with $Q=0.8$
((a) in Fig. 3). } \label{fig4}
\end{figure}
However, this is equivalent to Eq. (\ref{sol1}), which is one of
the attractor equations in the Sen's entropy function approach. As
is shown in Fig. 4, there exits a point of $V(\phi_0)=0$ if $Q \le
Q_c=1.325$, where $Q_c$ corresponds to the critical point
$(\tilde{x}_c, \tilde{q}_c)$. Hence, $V(\phi_0)=0$ is simply
another representation to the attractor equation (\ref{sol1}). On
the other hand, one may expect that the solution to Eq.
(\ref{tee34}) gives us some information on the location of the
degenerate horizon. It can be rewritten as
\begin{equation}
 U''(r)|_{r=r_e}=\frac{2Q^2}{16\phi^2_0}
\frac{1}{\cosh^{2}(\frac{Q^2}{4M\sqrt{\phi_0}})}
-\frac{Q^4}{32M\phi^{5/2}_0}
\frac{\sinh(\frac{Q^2}{4M\sqrt{\phi_0}})}{\cosh^{3}(\frac{Q^2}{4M\sqrt{\phi_0}})}
\label{tee45},
\end{equation}
which is unfortunately nothing but another attractor equation
(\ref{sol2}). Hence, it seems difficult to determine the location
of the degenerate horizon of an extremal RBH using the
conventional attractor equations of (\ref{tee44}) and
(\ref{tee45}).

In order to find the location of the degenerate horizon, we have to
find the general solution to  Eqs. (\ref{tee33}) and (\ref{tee34})
by choosing a conformal gauge of $g_{tx}=0$ as~\cite{GKL}
\begin{eqnarray}
\frac{d\phi}{dx}&=&2(J(\phi)-{\cal C}), \\
 ds^2&=&-(J(\phi)-{\cal C})dt^2+\frac{dx^2}{J(\phi)-{\cal C}},
\end{eqnarray}
where the 2D mass function $J(\phi)$ is given by
\begin{equation}
J(\phi)=\int^{\phi}V(\tilde{\phi})d\tilde{\phi}=\sqrt{\phi}+M
\tanh\Big(\frac{Q^2}{4M\sqrt{\phi}}\Big).
\end{equation}
Here ${\cal C}$ is a coordinate-invariant constant of the
integration, which is identified with the mass $M$ of the extremal
black hole. A necessary condition that a 2D dilaton gravity admits
an extremal RBH is that there exists at least one curve of
$\phi=\phi_0={\rm const}$ such that $J(\phi)=M$. Actually, we have
an important relation between the 4D metric function $U(r)$ and 2D
mass function $J(\phi)$ as
\begin{equation}
 \sqrt{\phi}U(r(\phi))=J(\phi)-M
\end{equation}
with $r=2\sqrt{\phi}$. In addition, $J(\phi)$ should be monotonic
in a neighborhood of $\phi_0$ with the attractor conditions
$J'(\phi_0)=V(\phi_0)=0$ and $J''(\phi_0)=V'(\phi_0)\not=0$ to
have the extremal black hole. These correspond to the attractor
conditions in Eq. (\ref{attcon}). First, $J(\phi)=M$ determines
the horizons $r=r_\pm$
\begin{equation} \label{EHJT}
U(\phi_\pm)=1-\frac{M}{\sqrt{\phi_\pm}} \Big[1-\tanh
\Big(\frac{Q^2}{4M\sqrt{\phi_\pm}}\Big)\Big]=0 \to U(r_\pm)=0.
\end{equation}
Considering the connection of $\phi_0=\frac{1}{4}r_e^2$, the
attractor  conditions of $J'(\phi_0)=0$ and $J''(\phi_0)\not=0$
implies the extremal configuraion
\begin{equation}  \label{cforebh}
U'(r_e)=0, ~~U''(r_e)\not=0.
\end{equation}
Then, combining Eq. (\ref{EHJT}) with Eq. (\ref{cforebh}) leads to
the condition for finding the degenerate horizon $r=r_e$.
Following Sec. 4, for $Q_e=Mq_e= 2 \sqrt{w_0}M$, we find the
location of the degenerate horizon, $r_e=Mx_e=4Mw_0/(1+w_0)$.
Here, we have the $AdS_2$ spacetime with negative constant
curvature
\begin{equation}
\bar{R}_2|_{r=r_e}=-\frac{2h}{\sqrt{\phi_0}}=
-\frac{1}{\sqrt{\phi_0}}U''(r_e)=-\frac{(1+\omega_0)^4}{16M^3\omega^3_0}.
\end{equation}
The generalized entropy function takes the form
\begin{equation}
\bar{F}^{RBH}(\phi_0)= - \sqrt{\phi_0} U''(r_e).
\end{equation}
Finally, for the magnetically charged extremal RBH, the desired
Bekenstein-Hawking entropy is given by
\begin{equation}
\bar{S}^{RBH}_{BH}=-\frac{4\pi\sqrt{\phi_0}}{U''(r_e)}\bar{F}^{RBH}(\phi_0)
=4\pi\phi_0=\pi r_e^2.
\end{equation}
Given $\tilde{x}=x_e$, $\tilde{q}=q_e$, this entropy can be
exactly recovered from Eq. (\ref{sen1}) in the entropy function
approach as
\begin{equation}
{\cal F}(x_e,q_e)=\pi M^2
q_e^2\cosh^{-2}\left(\frac{q^2_e}{2x_e}\right)=\pi r^2_e=\pi
M^2x^2_e.
\end{equation}
 Here, we also
note that the $AdS_2$ curvature $\bar{R}_2(U''(r_e))$ and
$\sqrt{\phi_0}$ are irrelevant to determining the entropy of the
extremal RBH. Furthermore, we confirm that the entropy is invariant
under the conformal transformation because $\sqrt{\phi_0}$ is a
conformal factor~\cite{CC}.

\section{Discussions}

We have discussed a magnetically charged  RBH  derived from  the
coupled action of Einstein gravity and nonlinear electrodynamics.
Although the action is simple, it is very interesting to investigate
its extremal black hole because its action is nonlinear on the
Maxwell side.  This black hole solution is parameterized by the ADM
mass $M$ and magnetic charge $Q$, while the free parameter
$a=Q^{3/2}/2M$ is adjusted to make the resultant line element
regular at the center. It turned out that the entropy function
approach does not lead to a correct entropy of the
Bekenstein-Hawking entropy.  This is mainly because the magnetically
charged extremal RBH is obtained  from the coupled system of
Einstein gravity and nonlinear electrodynamics. In the limit of $a
\to 0$ (Einstein-Maxwell theory), one finds the condition of $M=Q$
for the RN black hole. In this case, all approaches mentioned by
this work provide the Bekenstein-Hawking entropy $S_{BH}^{RN}=\pi
Q^2$ because of its linearity $\tilde{x}=\tilde{q}$, where
$r=M\tilde{x}$ and $Q=M\tilde{q}$. For $a=Q^{3/2}/2M$ case of the
extremal RBH, there is no linearity between $\tilde{x}$ and
$\tilde{q}$ and instead, their connection is determined by the
nonlinearity of
$\tilde{x}=\tilde{q}\cosh^{-1}(\tilde{q}^2/2\tilde{x})$ in Eq.
(\ref{sol1}). It follows that the entropy function approach is
sensitive to whether the nature of the central region of the black
hole is regular or singular .

The two  attractor equations in Eqs. (\ref{e33}) and (\ref{e332})
are not enough to determine the entropy of the extremal RBH
because we have a lot of solutions satisfying these same attractor
equations in Fig. 3. That is, Eq. (\ref{e33}) of
$\tilde{x}=\tilde{q}\cosh^{-1}(\tilde{q}^2/2\tilde{x})$ does not
imply the condition for determining the degenerate horizon of
$U(x)=U'(x)=0,~U''(x)\not=0$. Solving the Einstein equations in
the near horizon geometry is not sufficient to obtain the entropy
of the extremal RBH.  Hence, to find the correct form of the
entropy of extremal black hole, we introduce the generalized
entropy formula combined with a 2D dilaton gravity. In this case,
the  new attractor equations are given by Eqs. (\ref{tee33}) and
(\ref{tee34}), which contain full information on the location of
the degenerate horizon. Using the 2D dilation gravity approach,
the new attractor equations provide the condition of (\ref{EHJT})
and (\ref{cforebh}) for determining the location of the degenerate
horizon. Also we check that Eq. (\ref{e33}) is satisfied with
$\tilde{x}=x_e$ and $\tilde{q}=q_e$, corresponding to $\bullet$ in
Fig. 3.

At this stage, we would like to mention the difference between the
RN and RBH  black holes in obtaining  entropies of their extremal
black boles. In the case of the RN black hole, the attractor
equation of $\tilde{x}=\tilde{q}(r=Q)$ with the free parameter $M$
is enough to determine the Bekenstein-Hawking entropy as the
extremal black hole entropy. This means that the extremal
condition of $\tilde{x}=\tilde{q}=1(r=M=Q)$ is not necessary for
finding the extremal entropy. However, for the RBH with the free
parameter $M$, we have to know the extremal position of $x_e$ and
the charge $q_e$ exactly to obtain the entropy because the
attractor equation of
$\tilde{x}=\tilde{q}\cosh^{-1}(\tilde{q}^2/2\tilde{x})$ is
nonlinear.

In this work, we have succeeded to find  the entropy of the
extremal RBH by using the 2D dilation gravity approach. This
approach provides the location of horizon with attractor
conditions for degenerate horizon, which are
$U(x_e)=U'(x_e)=0,U''(x_e) \not=0$. We stress that this is not
possible if one does not use the dilaton gravity approach known as
the $s$-wave approximation of 4D gravity theory. Using Sen's
entropy function approach, one can get $U'(x_e)=0$ and $U''(x_e)
\not=0$ partly. In other words, although Sen's entropy function
approach is known to provide Einstein equation in the near horizon
geometry of $AdS_2$ spacetime as the attractor equations, this
does not work for the regular black hole. Formally, we have to use
full Einstein equations to find the entropy of  extremal regular
black hole. However, noting that the $s$-wave approximation
preserves the attractor conditions, we have used the 2D dilaton
gravity approach to find the location of the degenerate horizon
for simplicity, instead of solving the full Einstein equations.

In conclusion, we have successfully obtained the entropy of an
extremal regular black hole from the generalized entropy formula
based on Wald's Noether charge formalism combined with the 2D
dilaton gravity approach.

\vspace{0.5cm}

\medskip
\section*{Acknowledgments}
We would like to thank L.-M. Cao for useful discussions. Two of us
(Y. S. Myung and Y.-J. Park) were supported by the Science Research
Center Program of the Korea Science and Engineering Foundation
through the Center for Quantum Spacetime of Sogang University with
grant number R11-2005-021. Y. S. Myung  was also in part supported
by the Korea Research Foundation (KRF-2006-311-C00249) funded by the
Korea Government (MOEHRD). Y.-W. Kim was supported by the Korea
Research Foundation Grant funded by Korea Government (MOEHRD):
KRF-2007-359-C00007.

\end{document}